\documentstyle[psfig,conf_iap,10pt]{article}
\def\etal{{\em et al.}\ }
\def\deg{\hbox{$^\circ$}}
\def\arcmin{\hbox{$^\prime$}}
\def\arcsec{\hbox{$^{\prime\prime}$}}
\def\micron{\hbox{$\mu$m\ }}

\begin{document}
\heading{%
%
The European Large Area {\em ISO} Survey: ELAIS
%
} 
\par\medskip\noindent
\author{%
Seb Oliver$^{1}$, 
M. Rowan-Robinson,
C. Cesarsky,
L. Danese,
A. Franceschini,
R. Genzel,
A. Lawrence,
D. Lemke,
R. McMahon,
G. Miley,
J-L. Puget,
B. Rocca-Volmerange,
P. Ciliegi,
A. Efstathiou,
C. Gruppioni,
P. H\'eraudeau,
S. Serjeant,
C. Surace
and ELAIS consortium

}
\address{%
Imperial College of Science Technology and Medicine,
	   Astrophysics Group,
	   Blackett Laboratory,
	   Prince Consort Rd.,   London,
	   SW2 1BZ, 
	   s.oliver@ic.ac.uk}
%
\address
{To appear in  `Wide-field surveys in cosmology', Proc. XIV IAP meeting}


%

\begin{abstract}

The European Large Area {\em ISO} Survey (ELAIS) has surveyed  $\sim 12$ square
degrees of the sky at 15\micron and 90\micron and subsets of this
area at 6.75\micron and 175\micron using the  Infrared Space Observatory 
({\em ISO}).  This project was the largest single open time programme executed
by {\em ISO}, taking 375 hours of data.
A preliminary catalogue of more than 1000 galaxies has been
produced.   In this talk we describe the goals of the project 
and present a  provisional number count analysis at 15 \micron.

\section{Introduction}

The
European Large Area {\em ISO} Survey (ELAIS) is 
a collaboration involving 19 European institutes (in
addition to the authors and others at their institutes
the following people and others at their institutes are involved
I. Gonzalez-Serrano,
E. Kontizas,
K. Mandolesi,
J. Masegosa, 
K. Mattila,
H. Norgaard-Nielsen,
I. Perez-Fournon,
M. Ward)
and is the
largest open time project being undertaken by {\em ISO}.

\end{abstract}
\section{Key Scientific Goals}

In this section we highlight some of the principal 
scientific motivations for this programme, though naturally there
are many other goals which we have not space discuss.


The main extra-galactic population detected by IRAS was galaxies with
high rates of star formation. These objects are now known
to evolve with a strength comparable to AGN. 
 The sensitivity of {\em ISO} 
allow us to detect these objects at much higher redshifts and thus
obtain greater understanding of the cosmological evolution of star
formation.  This is directly complimentary to studies of
star-formation history in the optical UV. 
Comparison of the star-formation rates determined in the FIR
with that determined from the UV will give a direct estimate of the
importance of dust obscuration, vitally important for models of
cosmic evolution.


If elliptical galaxies underwent a massive burst of star-formation
between $2<z<5$, they would be observable in the far infrared and may
look like F10214.  This survey will provide a powerful
discrimination between such a top-down scenario and any hierarchical
bottom-up merging model whose components are
individually too faint to detect.


Unified models of AGN suggest that the central engine is surrounded by
a dusty torus.  The far infrared emission from the torus  is much
less sensitive to the viewing angle than the optical.  Thus a far
infrared selected sample of AGN will place
important constraints on unification schemes.


IRAS uncovered a population with enormous far infrared luminosities,
$L_{\rm FIR} $ $> 10^{12}L\odot$.    While most of
these objects appear to have an AGN it is argued that star formation
could provide most of the energy.  Interestingly, most of these
objects appear to be in interacting systems, suggesting a
triggering mechanism.  Exploration of this population to higher redshift
will have particular significance for models of AGN/galaxy evolution.

\section{The ELAIS Survey}

The {\em ISO} observations took place between 12th March 1996 and 8th April 1998,
shortly before the Helium boil off.  The total survey area covered in
blank fields is 6, 11, 12, \& 1 sq. deg. at 6.75, 15, 90 and 175 \micron,
an additional 2 sq. deg. within the ELAIS survey areas has been surveyed
by the FIRBACK team \cite{dc98}.  Table \ref{areas} specifies the main
survey areas.

\begin{center}

\begin{table}
\caption{Summary of ELAIS Survey Areas.
These areas were selected primarily for having low
Cirrus contamination, specifically $I_{100}<1.5$MJy/sr from
the IRAS maps of Rowan-Robinson \etal (1991).
For N1-3, S1 and all X areas we also restricted ourselves to regions of
high visibility $>25$\% over the mission lifetime.  For low Zodiacal 
background we
required  $|\beta|>40$ and to avoid saturation of
the CAM detectors we had to avoid any bright IRAS 12$\mu$m sources.
The 6 additional small ($0.3\deg \times 0.3\deg$) rasters (X1-6)
were centred on well studied areas of the
sky or high-$z$ objects}
\label{areas}

\begin{tabular}{lrrrrrc}
Name &
\multicolumn{2}{c}{Nominal Coordinates} &
M & N & ROLL &
 $\langle I_{100}\rangle$ \\

  & \multicolumn{2}{c}{J2000}
 & /\deg & /\deg & /\deg
 & \begin{small}$/{\rm MJy sr}^{-1}$ \end{small} \\
\\
N1&$16^h10^m01^s$&$+54\deg30\arcmin36\arcsec$&
 2.0 & 1.3 & 20 & 1.2 \\
N2 &$16^h36^m58^s$&$+41\deg15\arcmin43\arcsec$&
 2.0 & 1.3 & 30 &  1.1  \\
N3 &$14^h29^m06^s$&$+33\deg06\arcmin00\arcsec$&
 2.0 & 1.3 & 330 &  0.9 \\
S1 &$00^h34^m44^s$&$-43\deg28\arcmin12\arcsec$&
 2.0 & 2.0 & 20 &  1.1 \\

S2 &$05^h02^m24^s$&$-30\deg35\arcmin55\arcsec$&
0.3 & 0.3 & 290 & 1.1 \\




 \end{tabular}

\end{table}
\end{center}

\section{Comparison with other {\em ISO} surveys}

 {\em ISO} carried out a variety of surveys exploring the available 
parameter space of depth and area.  Table \ref{surveys} summaries
the main extra-galactic blank-field surveys.

\begin{table}
\begin{center}
\caption{Field Surveys with {\em ISO} \label{surveys}}
\begin{tabular}{lllll}
\hline
Survey Name  & [e.g. ref] & Wavelength & Integration & Area\\
             &    &   $/\mu$m  &   $/$s      & $/{\rm sq deg}$\\
\\
PHT Serendipity Survey & \cite{sb96}     & 175         & 0.5            & 7000 \\
CAM Parallel Mode & \cite{rs96}       & 7           & 150            & 33 \\
ELAIS         &     & 7,15,90,175 & 40, 40, 24, 128& 6, 11, 12,1\\
CAM Shallow   &   \cite{de98}         & 15          & 180            & 1.3  \\
FIR Back      &    \cite{dc98} & 175         & 256, 128       & 1, 3  \\
IR Back       &     \cite{km99}         & 90, 135,180 & 23, 27, 27     & 1, 1, 1 \\
SA 57         & \cite{hunn96} & 60, 90      & 150, 50        & 0.42,0.42  \\
CAM Deep      &     \cite{de98}    & 7, 15, 90   & 800, 990, 144  & 0.28, 0.28, 0.28\\
Comet fields       & \cite{dc99} & 12          & 302            & 0.11\\
CFRS          & \cite{fham98}                    & 7,15,60,90  & 720, 1000, 3000,3000 & 0.067.0.067.0.067,0.067\\
CAM Ultra-Deep&       \cite{de98}         & 7           & 3520           & 0.013 \\
ISOHDF South  & \cite{sjo99}           & 7, 15       &$>6400, >6400$  & 4.7e-3, 4.7e-3 \\
Deep SSA13    & \cite{ytan97}                & 7           & 34000          & 2.5e-3\\
Deep Lockman  &  \cite{kkawara98}      & 7, 90, 175  & 44640, 48, 128 & 2.5e-3, 1.2 , 1 \\
ISOHDF North  & \cite{sbgs97}        & 7, 15       & 12800, 6400    & 1.4e-3, 4.2e-3 \\
\hline
\end{tabular}
\end{center}
\end{table}

\section{Provisional Number Counts}

\begin{figure}
\centerline{\vbox{
\psfig{figure=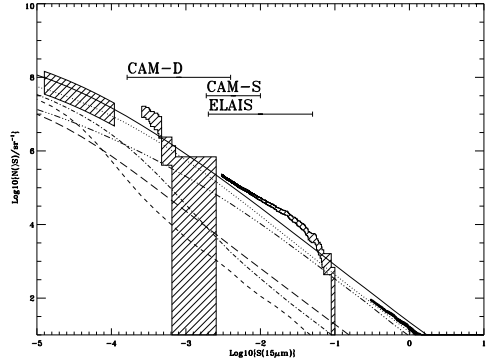,height=5.cm,angle=90}
}}
\caption[]{Provisional 15 \micron integral source 
counts from the ELAIS survey,
together counts from {\em ISO HDF} (\cite{sjo97})
with {\em IRAS} 12 $\mu$m counts (thick line) at bright end, 
(Rush {\em et al.} 1993 {\em IRAS} data shifted to 15 $\mu$m using 
cirrus spectrum).  Models are
from  \cite{af94}: 
all components solid;
spiral galaxies dotted;
star-bursts dash-dot-dot-dot;
s0 dot-dash;
AGN-dash
Elliptical galaxies-short dashes.
Faint end constraints (hatched) come from the {\em ISO HDF} $P(D)$ analysis.
The depths probed by some other {\em ISO} surveys are indicated 
CAM-D \& CAM-S \cite{de97}.
}\label{af_15}

\end{figure}

Source catalogues have been extracted from the ELAIS data at all 
wavelengths and a very provisional source count analysis has been
performed at 15 \micron.  The results of this are shown in Figure
\ref{af_15}.  As can been seen these counts confirm the strong
evolution detected in the {\em ISO HDF} analysis \cite{sjo97}.
A more detailed analysis of the counts will be presented shortly 
\cite{sbgs99}

%

\section{Follow-up}

An extensive follow-up programme is being undertaken,
including measurements at all wavelengths from X-ray to radio \cite{pc98}.  This
programme will provide essential information for identifying the
types of objects detected in the infrared, their luminosities,
energy budgets and other detailed properties.  As well as
studying the properties of the objects detected by {\em ISO} a number
of the follow-up surveys will provide independent source lists
which will be extremely valuable in their own right, not least because
we can investigate why some objects emit in the infrared and others
do not.

\acknowledgements{This paper is based on observations with {\em ISO}, an ESA project, with
instruments funded by ESA Member States (especially the PI countries:
France, Germany, the Netherlands and the United Kingdom) and with
participation of ISAS and NASA.
This work was in part supported by PPARC grant no.  GR/K98728
and  EC Network is FMRX-CT96-0068
 }


%
\vfill
\end{document}